\documentclass[aip,prl,graphicx,reprint]{revtex4-1}
\usepackage{graphicx}
\usepackage{mathtools}
\usepackage{xcolor}

\begin{document}
\title[An Haloscope Amplification Chain based on a Travelling Wave Parametric Amplifier]{An Haloscope Amplification Chain based on a Travelling Wave Parametric Amplifier}

\author{Caterina Braggio}
\affiliation{Dip. di Fisica e Astronomia, Universit\`a di Padova, 35100 Padova, Italy}
\affiliation{INFN - Sezione di Padova, 35100 Padova, Italy}
\author{Giulio Cappelli}
\affiliation{Univ. Grenoble Alpes, CNRS, Grenoble INP, Institut N\'eel, 38000 Grenoble, France}
\author{Giovanni Carugno}
\affiliation{INFN - Sezione di Padova, 35100 Padova, Italy}
\author{Nicol\`o Crescini}
\affiliation{Univ. Grenoble Alpes, CNRS, Grenoble INP, Institut N\'eel, 38000 Grenoble, France}
\author{Raffaele Di Vora}
\affiliation{INFN - Sezione di Padova, 35100 Padova, Italy}
\affiliation{Dip. di Scienze Fisiche, della Terra e dell’Ambiente, Universit\`a di Siena, Italy}
\author{Martina Esposito}
\affiliation{Univ. Grenoble Alpes, CNRS, Grenoble INP, Institut N\'eel, 38000 Grenoble, France}
\affiliation{CNR-SPIN Complesso di Monte S. Angelo, via Cintia, Napoli, 80126, Italy}
\author{Antonello Ortolan}
\affiliation{INFN - Laboratori Nazionali di Legnaro, 35020 Legnaro, Padova, Italy}
\author{Luca Planat}
\affiliation{Univ. Grenoble Alpes, CNRS, Grenoble INP, Institut N\'eel, 38000 Grenoble, France}
\author{Arpit Ranadive} 
\affiliation{Univ. Grenoble Alpes, CNRS, Grenoble INP, Institut N\'eel, 38000 Grenoble, France}
\author{Nicolas Roch}
\affiliation{Univ. Grenoble Alpes, CNRS, Grenoble INP, Institut N\'eel, 38000 Grenoble, France}
\author{Giuseppe Ruoso}
\email[Corresponding author: ]{Giuseppe.Ruoso@lnl.infn.it}
\affiliation{INFN - Laboratori Nazionali di Legnaro, 35020 Legnaro, Padova, Italy}

\begin{abstract}
In this paper we will describe the  characterisation of a rf amplification chain based on a travelling wave parametric amplifier (TWPA). The detection chain is meant to be used for dark matter axion searches and thus it is mounted coupled to a high Q microwave resonant cavity. A system noise temperature $T_{\rm sys} = (3.3 \pm 0.1$)  K has been measured at a frequency of 10.77 GHz, using a novel scheme allowing  measurement of $T_{\rm sys} $ exactly at the cavity output port.
\end{abstract}
\today
\maketitle



\section{Introduction}

\subsection{Axions and haloscopes}

In the attempt to solve the so called strong CP problem of quantum chromodynamics (QCD), in the seventies a new hypothetical particle came into play: the axion\cite{pq,weinberg1978new,wilczek1978problem}. The axion is an extremely light weakly interacting pseudoscalar boson, described by two main classes of models: KSVZ \cite{PhysRevLett.43.103,SHIFMAN1980493} and DFSZ \cite{DINE1981199,Zhitnitsky:1980he}, generically indicated as QCD-axion models. It was soon realized that the axion could be the main component of galactic dark matter halos, an hypothesis supported by the presence of various mechanisms for axion production in the early Universe\cite{PRESKILL1983127,Sikivie2008,Duffy_2009,MARSH20161}. Astrophysical and cosmological constraints \cite{RAFFELT19901,turner1990windows}, as well as lattice QCD calculations of the DM density \cite{borsanyi2016calculation,berkowitz2015lattice,Buschmann2022}, provide a preferred axion mass window around tens of $\mu$eV.

Experimental searches of the axion are carried out in a wide variety of experiments \cite{Irastorza:2018dyq}, among which the most sensitive ones are the haloscopes: apparata trying to detect the axion forming the Milky Way halo. The detection principle was outlined by Sikivie in 1983 \cite{PhysRevLett.51.1415}: axion induced microwave photons are produced in a resonant cavity immersed in a static magnetic field through the inverse Primakoff effect. Ultra low noise microwave detection chains then look for excess radio frequency power.
Sikivie's type haloscopes allowed to exclude QCD-axions as the main DM component for axion masses $m_a$ between 1.91 and 4.2\,\,$\mu$eV \cite{PhysRevLett.120.151301,PhysRevLett.104.041301,PhysRevLett.124.101303, PhysRevLett.127.261803}, and, together with helioscopes \cite{Anastassopoulos2017}, are the only experiments which reached the QCD-axion parameter space. 


In a standard haloscope the detector is composed by a high quality factor $Q_0$ resonant cavity immersed in a static magnetic field $B_0$. The  signal is the power collected by an antenna coupled to a specific cavity mode with coupling $\beta$.
The expected axion power would be (for an antenna coupling $\beta=2$):
\begin{eqnarray}
P_{\rm a} =\, & 1.48 \cdot 10^{-24}\, {\rm W} 
\left( \frac{V \times C_{mnl} }{2.85 \cdot 10^{-5}\, {\rm m}^3} \right)
\left( \frac{B_0}{10\, {\rm T}} \right)^2 \times\\
\left( \frac{g_\gamma}{-0.97} \right)^2&
\left( \frac{\rho_a}{0.45\, {\rm GeV\, cm}^{-3}} \right)\nonumber
\left( \frac{\nu_c}{10\, {\rm GHz}} \right)
\left( \frac{Q_0}{100\,000} \right).
\end{eqnarray}

where $\rho_a=0.45$\,GeV/cm$^3$ is the local DM density, $g_{\gamma}$ is a model dependent parameter equal to $-0.97$ $(0.36)$ in the KSVZ (DFSZ) axion model, $C_{mnl}\simeq O(1)$ is a geometrical factor, depending on the cavity mode, 
$V$ is the volume of the cavity and $\nu_c$  its resonance frequency.
A  measurement at a fixed cavity frequency would be able to probe  axions with masses $m_a c^2 \simeq \hbar \nu_c$, and thus it is necessary to tune the cavity frequency to search for different mass values.
A practical parameter to characterize this system is  the scanning speed, i.e. the mass range that can be explored within a certain amount of time. 

The  sensitivity of axion search experiments is determined by the signal to noise ratio ${\rm SNR} \equiv {P_{\rm a}}/{\delta P_{\rm noise}}$. 
Following Ref. [\onlinecite{Lamoreaux:2013koa}] the expected noise is
\begin{equation}
	\delta P_{\rm noise} = k_B T_{\rm sys} \sqrt{\frac{\Delta\nu_a}{\Delta t}}
\end{equation}
where  $T_{\rm sys}$ is the equivalent system noise temperature, $\Delta t$ is the integration time and $k_B$ is the Boltzmann constant. $\Delta\nu_a \simeq \nu_c / Q_a$ is the intrinsic width of the axion signal, it is due to the velocity spread of dark matter particles in the halo and it can be shown that for the isothermal model an axion "quality factor" $Q_a=1\times 10^6$ should be used \cite{PhysRevD.42.3572}.
The general form of the scanning rate is derived as \cite{Kim_2020}
\begin{equation}
\frac{d\nu}{dt} =   \frac{1}{\rm SNR^2} \left(\frac{\beta P_a}{k_BT_{\rm sys}}\right)^2  \frac{1}{(1+\beta) Q_0}.
\label{eq:scan_rate_rev}
\end{equation}

 
It is then evident that reducing the system noise temperature is a main challenge  for this type of experiments.


\subsection{Amplification chains}

The choice of the amplification chain is closely related to the frequency at which the search for dark matter axion is conducted. A thorough discussion can be found in~[\onlinecite{Lamoreaux:2013koa}]. Most of running experiments are using a heterodyne or super-heterodyne detection chain with a linear amplifier as first stage. The linear amplifier is directly sensing the output of the antenna coupled to the cavity mode.  Recently a more advanced detection scheme has been used to take advantage of squeezing \cite{HAYSTAC:2020kwv}. 

The experiment QUAX \cite{PhysRevD.103.102004} is focusing its search efforts in a region around 10 GHz. In such a band commercially available amplifiers are based on High Electron Mobility Transistor (HEMT) with MMIC technology (monolithic microwave integrated circuits) showing an equivalent noise temperature of the order of 3.5 K \cite{8500350,9459460} at the temperature of 4.5 K. The system noise temperature is normally the sum of amplifier equivalent noise temperature and ambient temperature. 
Ambient temperature can be reduced by placing the apparatus inside an ultra cryogenic environment, however HEMTs equivalent noise temperature does not decrease significantly and they are normally not suitable to be used in such environments due to their large power dissipation.
Another solution for the first stage amplification is the use of Josephson Parametric Amplifiers (JPA) \cite{10.1063/1.1754906}, that can be operated at much lower temperature and indeed have already been used in haloscopes  \cite{Zhong_2018,PhysRevLett.124.171801,PhysRevLett.127.261803}. JPAs are quantum limited amplifiers, having the drawback of being resonant systems working only in a small frequency region and with little tuning: a 10 MHz gain profile can normally be tuned over a few hundreds of MHz \cite{PhysRevLett.108.147701}. This limitation has been overcome by Travelling Wave Parametric Amplifiers (TWPAs) consisting of nonresonant nonlinear transmission lines exhibiting amplification bandwidth up to few GHz \cite{9134828,PhysRevLett.60.764}. Recently such amplifiers have been used also for dark matter searches \cite{2021arXiv211010262B}.

In this paper we will describe an experimental test conducted on a haloscope amplification chain based on a TWPA realized using a reversed Kerr phase matching mechanism  \cite{ranadive2022}. The TWPA will be reading the power delivered by a microwave  cavity, resonating at frequencies typical for the QUAX axion haloscope.  The test has been conducted in the absence of the strong magnetic field which is used in axion searches, however the set-up is already equipped with a magnetic shielding and it will be soon incorporated in a complete haloscope. The present test is used to define the base working properties of the detection chain.

The precise measurement of the system noise temperature is of paramount importance in the assessment of the  results for an haloscope. The standard Y-factor method \cite{Pozar:882338}  normally uses a variable temperature source accessed by means of switches. Here we give details of a novel method that we  introduced previously \cite{PhysRevLett.124.171801}, with the advantage of avoiding the need for switches and capable of measuring the noise temperature exactly at the point of interest, namely at the haloscope cavity output.

\section{Experimental apparatus}
\label{experiment}
\begin{figure}[htb]
  \centering
     \includegraphics[width=.49\textwidth]{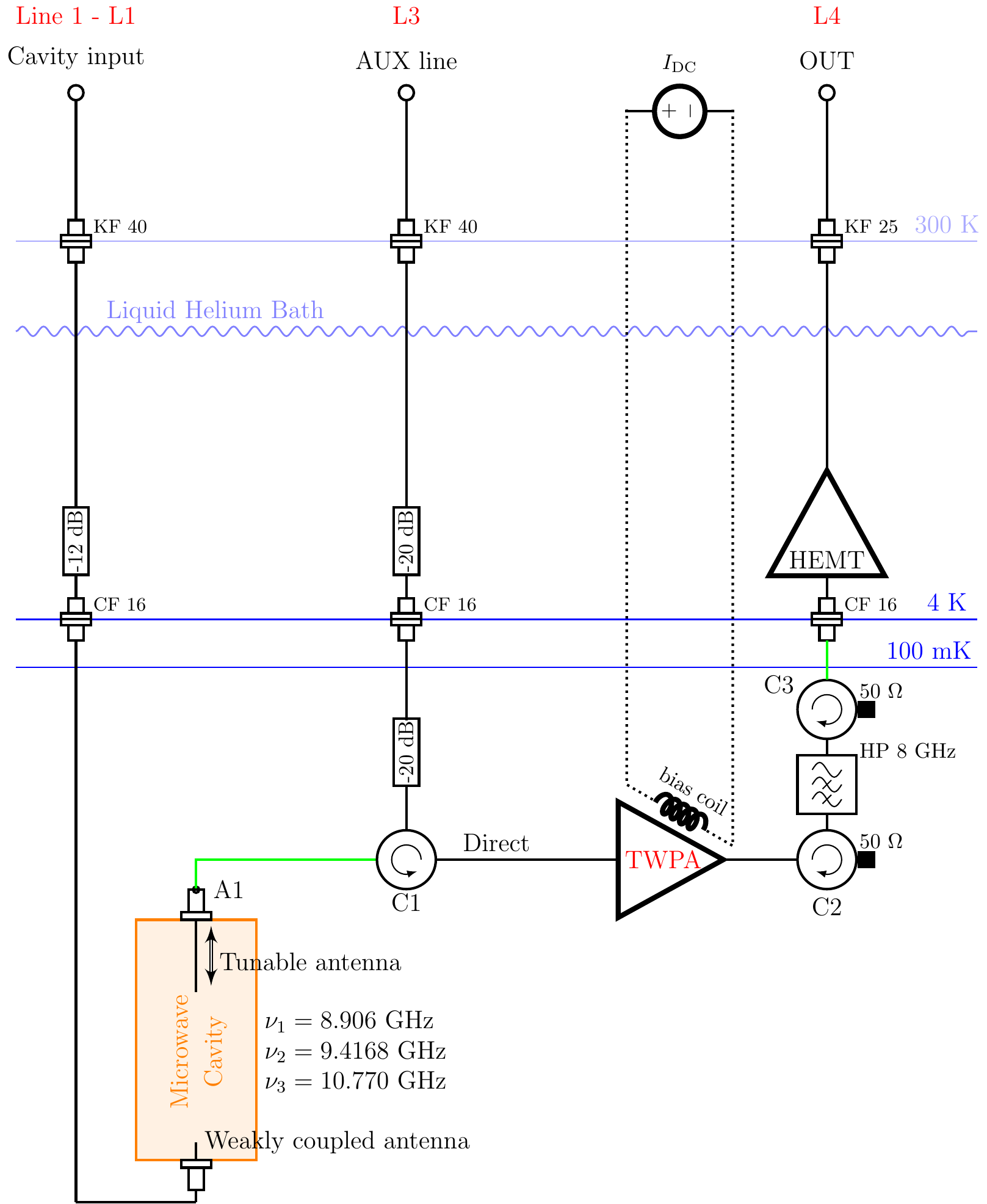}
\caption{\small Layout of cryogenic apparatus. Green cables are superconducting (NbTi). The point A1 is the reference point for the measurement of the system noise temperature. The connections between the TWPA and the two circulators C1 and C2 are done with direct SMA male to male adaptors. }
\label {fig:Apparatus}
\end{figure}

The schematic of the apparatus is reported in Fig.~\ref{fig:Apparatus}.

\begin{figure}[htb]
	\centering
	\includegraphics[width=.4\textwidth]{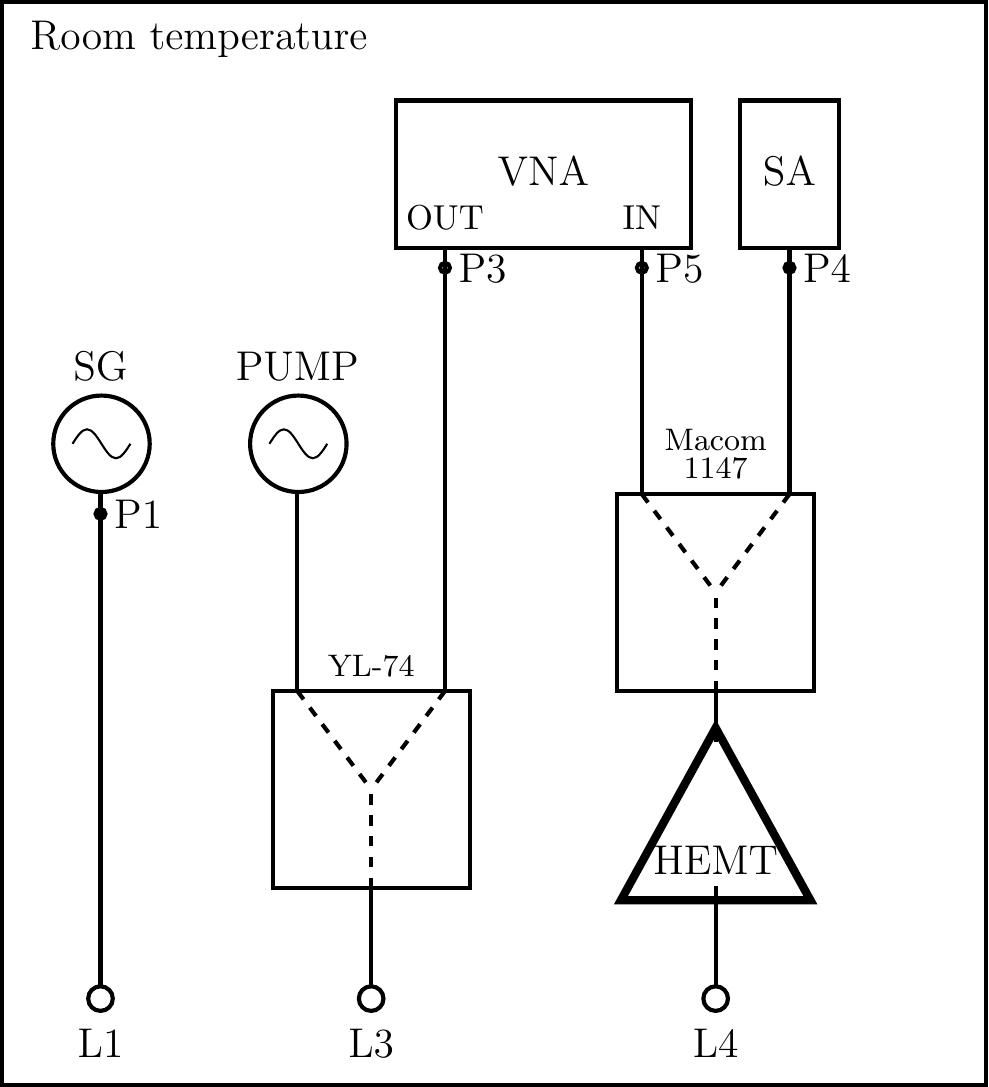}
\caption{\small Principle scheme of the room temperature radio frequency instrumentation. The points Pn are the reference points used in the measurements. SG = Signal generator; VNA = Vector Network Analyser; PUMP = Signal generator used as TWPA pump; SA = Spectrum Analyser. YL-74 and Macom 1147 are power splitters/combiners. L1, L3 and L4 are connected to the corresponding points of Figure \ref{fig:Apparatus}.}
\label {fig:rt}
\end{figure}

The element to be read by the TWPA is a cylindrical copper cavity of radius 12.88 mm and length 50 mm. Readout is performed by means of a dipole antenna whose coupling can be manually controlled with a mechanical feedthrough. The cavity relevant modes that have been used are the following: TM010 with resonant frequency $f_{\rm 010} = 8.91$ GHz, TM011 at  $f_{\rm 011} = 9.42$ GHz and TM012 at $f_{\rm 012} = 10.77$ GHz. The antenna output is fed onto a circulator (C1)  using a superconducting NbTi cable. C1 is  directly connected to the input of the TWPA, which serves as pre-amplifier of the system detection chain under test. Further amplification at a cryogenic stage is done using a low noise HEMT amplifier (Low Noise Factory model LNA4-16). In order to avoid back action noise from the HEMT, a pair of isolators (C2 and C3) and a 8 GHz High Pass filter are inserted between the TWPA and the HEMT. The output of the cryogenic HEMT is then delivered to the room temperature electronics (RTE) described in figure \ref{fig:rt}: we refer to this line as L4. Access to the microwave cavity is also possible by using a weakly coupled antenna, connected to the room temperature electronics through line L1. This line employs a 12 dB attenuator to avoid contribution of thermal inputs; moreover, the last part of the line is done using a cryogenic cable having low transmission (insertion loss about 16 dB) and low heat conductance. An auxiliary line, L3 in figure, is used for calibrations and to connect the RTE directly to the cavity tunable antenna by means of circulator C1. Again, line L1 is equipped with attenuators to avoid thermal inputs from stages at temperatures above the system base temperature. Finally, a dc current source is connected to a superconducting coil used to bias the TWPA.

The room temperature electronics employs two signal generators: following figure \ref{fig:rt}, SG is used to provide known signal as input to the system, PUMP is used to provide the pump signal to the parametric amplifier. Its output is summed up with the output of a VNA, which is used to produce transmissions spectra of the system. The output line L4 is amplified using a HEMT amplifier and can be read simultaneously by the VNA and by a spectrum analyser (SA). In Figure \ref{fig:rt} we labelled the points P1, P3, P4 and P5, which are used in the calibration procedure described below.

Since we plan to use this system inside an haloscope, it is necessary to shield the electronics from the presence of stray fields. In our haloscope \cite{PhysRevD.103.102004}, an 8 T magnetic field is acting on a microwave cavity, and a counterbias solenoid is used to reduce the field on the detection electronics. We have estimated a residual stray field of about {\color{black}10 mT}. To shield such field a hybrid box is encapsulating the two circulators C1 and C2 and the TWPA. This hybrid box is constituted by an external box of lead and an internal one of cryoperm. The box dimension is $35\times 65\times 210$ mm$^3$, with one small base opened to allow cabling, and is thermally anchored to the dilution unit  (DU) mixing chamber.
 


The cryogenic and vacuum system is composed of a cryostat and a $^3$He–$^4$He  wet dilution refrigerator. 
The cryostat is a cylindrical vessel of height 2300 mm, outer diameter 800 mm, inner diameter 500 mm (made by Precision Cryogenics System Inc). The dilution refrigerator is a refurbished unit (made by Leiden Cryogenics Inc.) previously installed in the gravitational wave bar antenna Auriga test facility \cite{Marin_2002}. Such dilution unit has a base temperature of 70 mK and cooling power of 1 mW at 120 mK.
The DU is decoupled from the gas handling system through a large concrete block sitting on the laboratory ground via a Sylomer carpet where the Still pumping line is passing. This assembly minimizes the acoustic vibration  induced on the TWPA,  which is  rigidly connected to the mixing chamber. 
Once the Helium cryostat has been filled up with liquid helium the DU column undergo a fast pre-cooling down to liquid-helium temperatures via helium gas exchange on the Inner Vacuum Chamber (IVC). This cooling down operation take almost 4 hours. When a temperature of 4 K has been reached, the pre-cooling phase ends, the inner space of the IVC is evacuated.  From that point on the dilution refrigerator takes over and the final cooling temperature of around 80 mK is attained after about 5 hours. No charcoal pump was present in the DU cooled system. A pressure of around 10$^{-7}$ mbar was monitored without pumping on the IVC room temperature side through all the experimental run.
Temperatures are measured with a set of different thermometers. Most of them are used to monitor the behaviour of the dilution unit. A RuO thermometer is used to monitor the cavity temperature. It is to be noted that the 20 dB attenuator close to C1 is thermally anchored to the mixing chamber.

\section{Measurements}
\label{Measurements}

\subsection{General parameters}

The first measurements to be performed are related to the optimisation of the TWPA working parameters \cite{ranadive2022}. This correspond to the choice of the bias current driving the superconducting coil and of the pump frequency and amplitude.  By measuring with the VNA the transmission spectra P3 -$>$ P5 while varying the dc current $I_{DC}$, it is possible to identify the working point of the TWPA as this where the S parameter S53 has a local minimum. Indeed, minima are found with a periodicity of about 4 mA. They are not symmetric around zero due to the presence of the magnetic field of the circulators C1 and C3, which are located close to the TWPA. Our minimum currrent working point have been selected as $I_{DC}=\{ -1.2, 2.7\}$ mA. The behaviour of the TWPA for either one of the two values of $I_{DC}$ is equal.

\begin{figure}[htb]
	\centering
	\includegraphics[width=.47\textwidth]{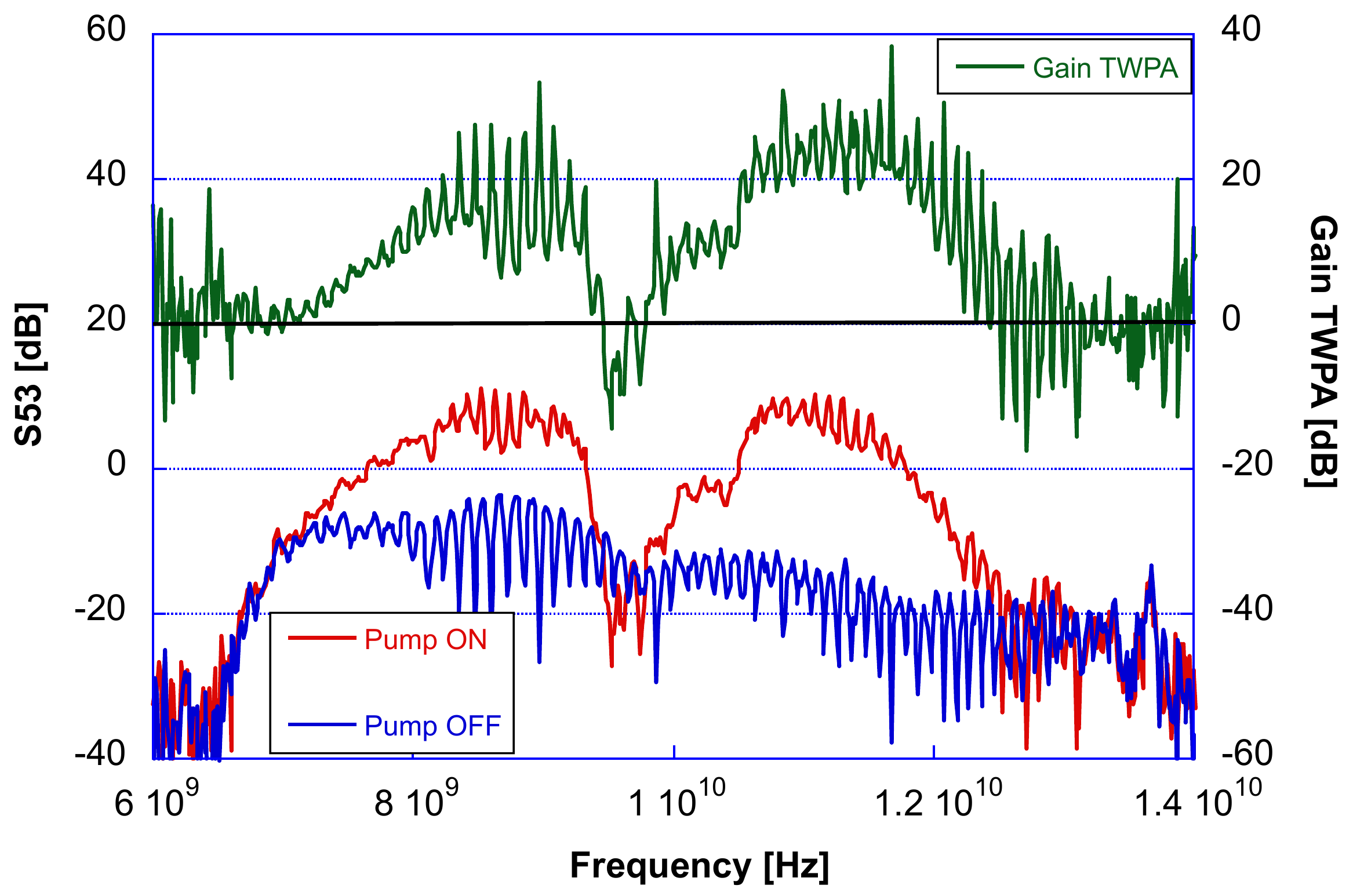}
\caption{\small Gain profile for the TWPA operated with a pump at the frequency of $f_P = 9.965$ GHz, with a biasing magnetic field produced by the current $I_{DC} = 2.7$ mA. A rough estimate of the gain is simply obtained by subtracting the S53 measure with pump OFF from that with pump ON.}
\label {fig:gain}
\end{figure}

Once the correct magnetic flux on the TWPA has been selected, the amplifier can be driven with a pump. Importantly, the TWPA used in this work does not rely on a dispersive feature for phase matching and can then be pumped at any desired frequency. Then the choice of the pump frequency and power is directly connected with the frequency of the signal we want to study. Among the available cavity resonances, we have selected the mode TM012 at $f_{\rm 012} = 10.77$ GHz. The antenna has been almost critically coupled to this mode, resulting in a mode width of about 1 MHz, with a corresponding loaded quality factor of about 10\,000. For this signal frequency a suitable pump frequency has been found as $f_P = 9.965$ GHz, with $I_{DC} = 2.7$ mA.  Figure \ref{fig:gain} shows the gain profile obtained with a power of $ -24.3$ dBm delivered from the signal generator PUMP.  The actual gain of the TWPA is estimated by dividing the S53 spectra obtained with the generator PUMP on and off. It has to be noted that this is an upper limit for the gain, since the real gain would be measured with respect to the situation where the TWPA is removed from the system and substituted with a quasi-lossless line. We could not do this in our set-up, since this would need the presence of a pair of switches, not available at the moment. The pump power level has been chosen by reducing of some fraction of a dB the value showing a saturation in the gain profile.

\subsection{Lines gains and Noise Temperature}

With the TWPA correctly in operation, it is then possible to complete the characterisation of the system by measuring the effective gain and noise temperature of the overall detection chain from the cavity output to the point P4. To this end, calibrated signal are injected and read along the different lines L1, L3 and L4, to allow the measurement of the transmission characteristic of all lines down to the common reference point given by the tunable antenna (point A1 of Figure \ref{fig:Apparatus}). We can define the following frequency dependent line gains:
\begin{description}
\item[$g_1(f)$]
from the point P1 to antenna A1 - bidirectional
\item[$g_3(f)$]
from the point P3 to antenna A1 - bidirectional 
\item[$g_4(f)$]
from antenna A1 to the point P4 (Complete detection chain)
\end{description}

Measurements are performed only at a cavity resonance, with the tunable antenna almost critically coupled. Only for $g_3$ a slightly detuned frequency is used, just off the cavity resonance.  

We then measure the following three transmission power spectra:

\begin{itemize}
\item S41: From P1 to P4 
\item S43: From P3 to P4 
\item S13: From P3 to P1
\end{itemize}

A calibrated signal generator provides as input a pure tone of different power levels $P_i$ and the output power levels $P_o$ are obtained with a spectrum analyser set to a specific resolution bandwidth $B$. 

\begin{equation}
P_o^{xy} = P_n + P_i \times G_{xy}, \,\,\,\,\, (xy)=\{41,43,13\}
\label{equa}
\end{equation}

where $P_n$ is the noise power at the spectrum analyser and $G_{xy}= g_x\times g_y$. The spectra $S41$ and $S43$ are those carrying the information on the noise level in the system: 
\begin{equation}
P_n = g_4 k_B T_{\rm sys} B + P_{\rm SA}^n
\label{equa2}
\end{equation} 

with $k_B$ the Boltzmann constant, $T_{\rm sys}$ the system noise temperature of the detection chain measured at the antenna point A1, and $P_{\rm SA}^n$ the  noise floor of the spectrum analyser. 

A linear  fit of the measured values for the three equations (\ref{equa}), i.e. with $(xy)=\{41,43,13\}$ allows to obtain the three line gains  $G_{xy}$ and $P_n$.
The value of $g_4$, together with $g_1$ and $g_3$, is then directly derived from $G_{xy}$ and using equation (\ref{equa2}) the resulting value of $T_{\rm sys}$ is calculated, characterising the noise performance of the detection chain.

\subsubsection{Systematics}

It is assumed that the cavity output line transmissivity shows no variation, due to non correct impedance matching, between the cavity mode frequency and the frequency at which S43 is measured. This is in general true for the slight shift used (order of a few MHz). For this reason the spectra $S$43 are measured in two positions, one above and one below the cavity resonance frequency. Another possible source of error is the leakage of circulator C1 from line L3 directly to line L4. From the specifications of the device and our measurements at room temperature we expect this leakage of the order of a percent. A more important source of error is the calibration of the generator providing the power input. We have checked the nominal power level of the instrument by comparison with other rf devices present in our laboratory: differences were below 0.1 dB, namely again at the percent level.

\begin{figure}[hb]
	\centering
	\includegraphics[width=.44\textwidth]{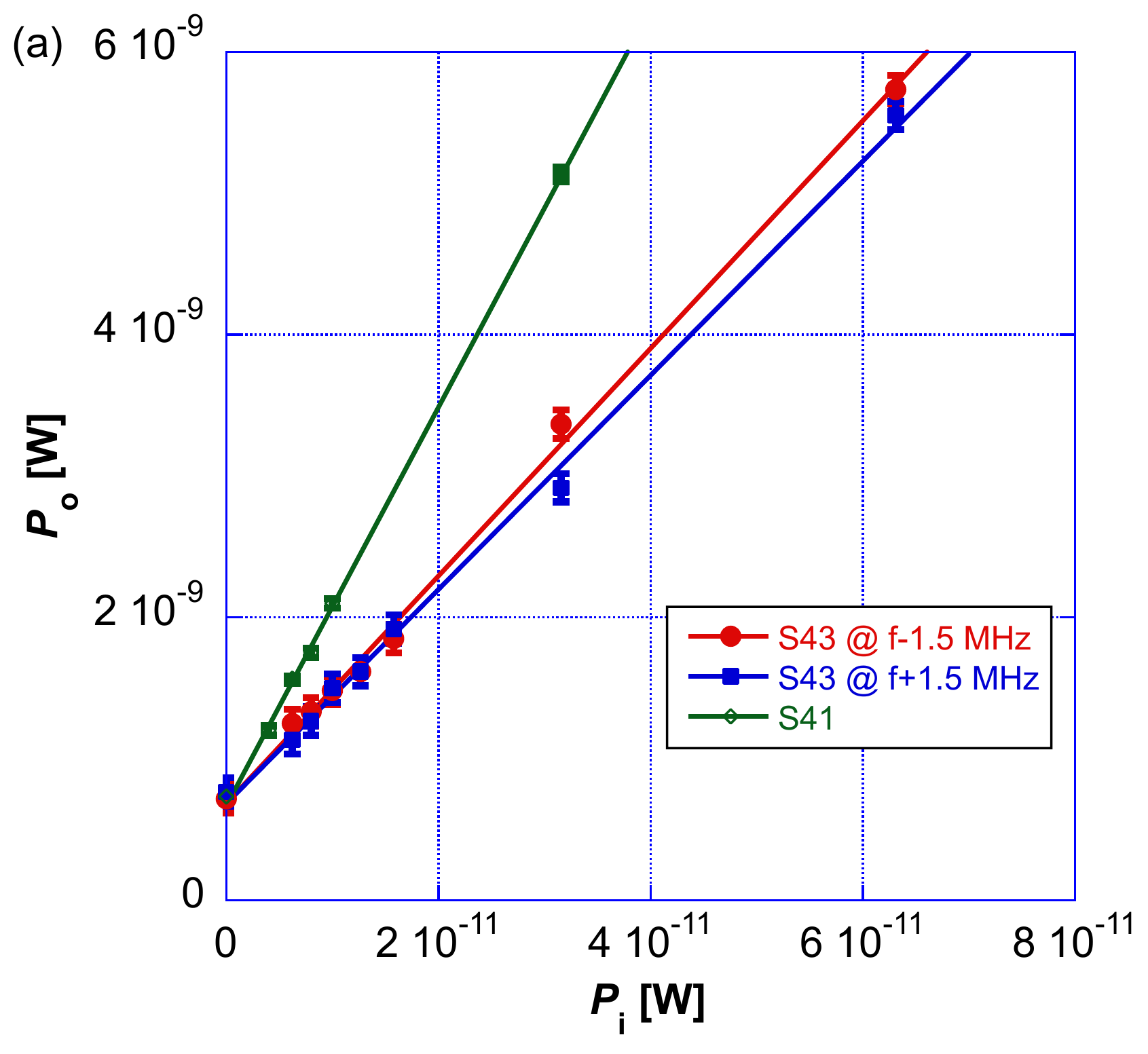}
	\includegraphics[width=.44\textwidth]{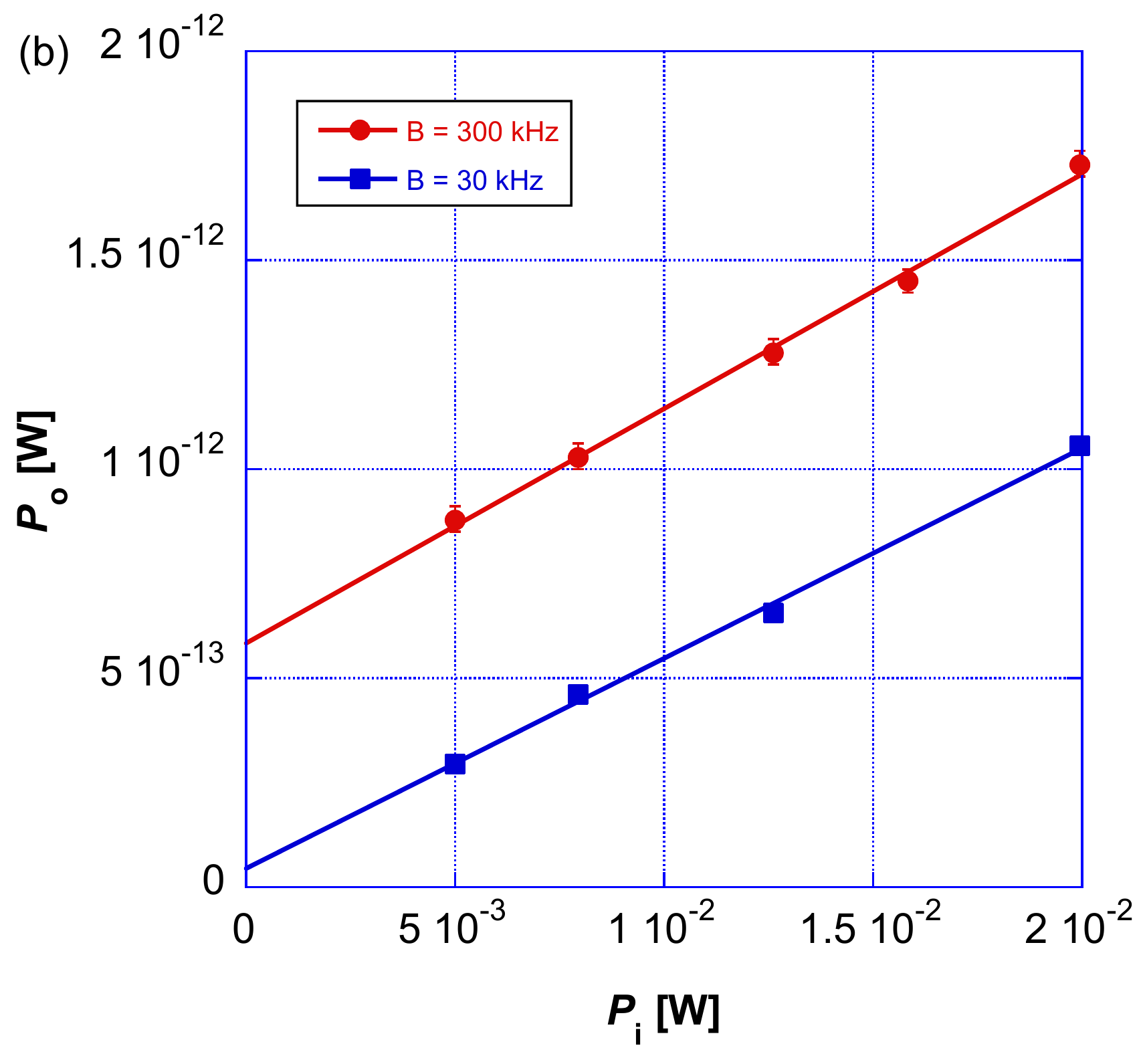}
\caption{\small (a) Measurements with different input powers for $S$43 and $S$41. The resolution bandwidth is $B=$ 1 MHz. The $S$43 measurements have been done at two frequencies, shifted $+ 1.5$ MHz and $- 1.5$ MHz from the resonance of the cavity. 
(b) Measurement for the quantity $S$13, performed with two bandwidths of 30 kHz and 300 kHz. For both panels least squares linear fits to the experimental points are shown.}
\label {fig:tn}
\end{figure}

\subsubsection{Results}

Figure \ref{fig:tn}a shows the measurement fo the quantities $S$43 and $S$41, having a resolution bandwidth $B = 1$ MHz. The noise of the spectrum analyser has been previously measured as $ P_{\rm SA}^n = 2 $ pW. For $S$43 two measurements have been done at two frequencies outside the cavity peak: each of them shifted 1.5 MHz above and below the cavity resonance value, respectively. 
For the case os $S$43 the obtained values are the average values of the two fits.
Figure \ref{fig:tn}b shows the measurement for the quantity $S$31, in this case two smaller bandwidth values (30 kHz and 300 kHz) have been used  since the output power level to be measured was very weak. Results of the fits are reported in the left column of Table \ref{t:results}. The errors are those obtained from the fit, each measured point has its own error given essentially by the number of averages taken for each point. The right column of the table shows the values of the parameters calculated  using the definition of  $G_{xy}$ and Equation   (\ref{equa2}).

\begin{table}[htp]
\caption{Results from the fits of Figure \ref{fig:tn} and calculated values for the measurement  parameters.}
\begin{center}
\begin{tabular}{|c|c|}
\hline
Parameters from fits & Calculated parameters \\
 \hline
$G_{41} =141 \pm 1$  & $g_1  = (9.8  \pm 0.7) \times 10^{-6}$ \\
\hline
$G_{43} = 78 \pm 2$ & $g_3  = (5.4  \pm 0.4) \times 10^{-6} $ \\
\hline
$G_{13} = (5.3 \pm 0.2) \times 10^{-11}$  & $g_4  = (1.44 \pm 0.07) \times 10^7$ \\
\hline
\hline
$P_n^{43}= (6.9 \pm 0.5)\times 10^{-10}$ W & $T_{sys}^{43} = 3.5 \pm 0.3$ K \\
\hline
$P_n^{41}= (6.7 \pm 0.2)\times 10^{-10}$ W & $T_{sys}^{41} = 3.4 \pm 0.2$ K \\
\hline
\end{tabular}
\end{center}
\label{t:results}
\end{table}%
\begin{figure}[htb]
	\centering
	\includegraphics[width=.46\textwidth]{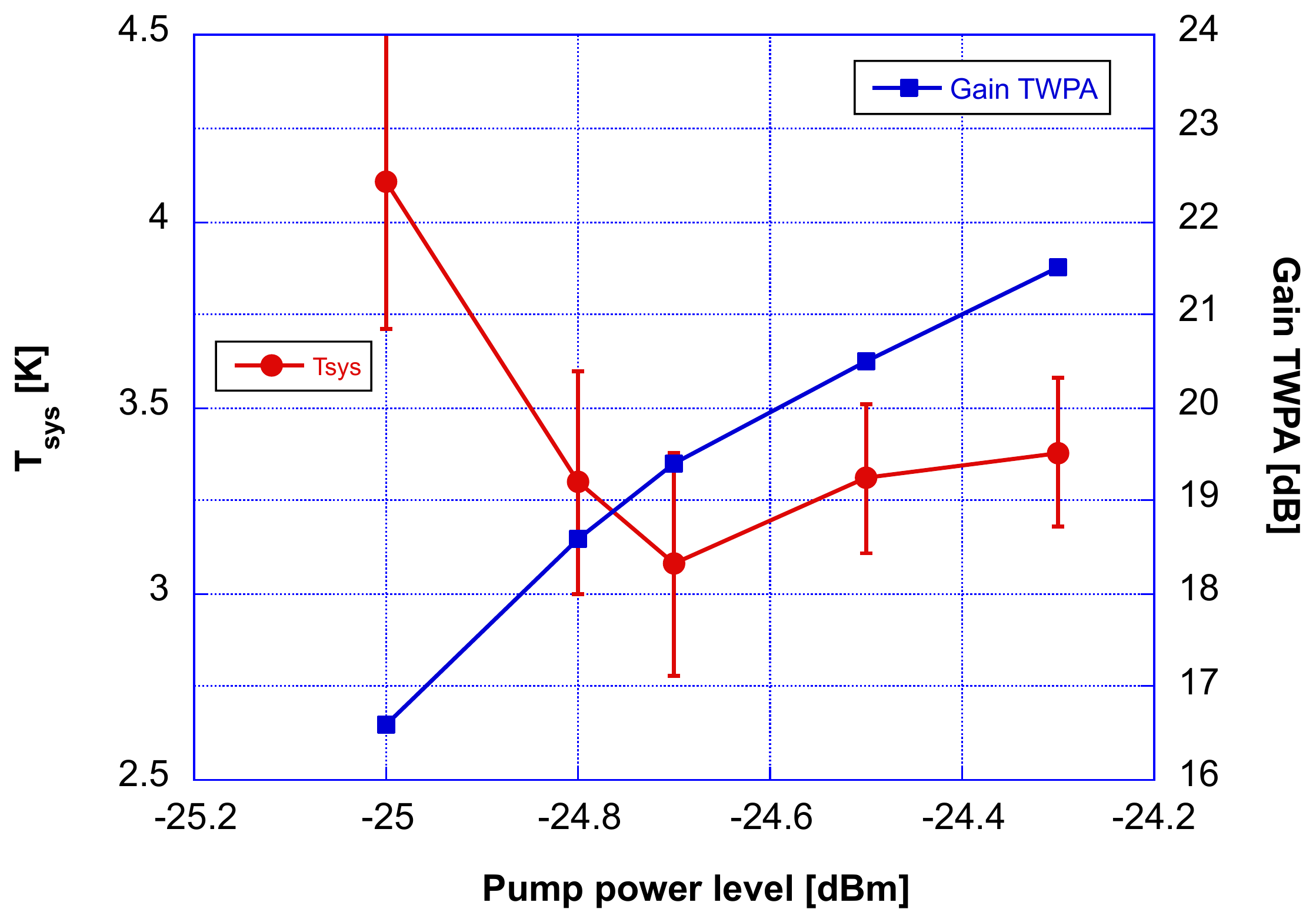}
\caption{\small System noise temperature and TWPA gain for different values of the pump power level. The pump frequency is $f_P = 9.965$ GHz,  with a biasing magnetic field produced by the current $I_{DC} = 2.7$ mA.}
\label {fig:vp}
\end{figure}

We have studied the variation of the system noise temperature for different gains of the TWPA, obtained by varying  the pump power level at the frequency $f_P = 9.965$ GHz. Results are shown in Figure~\ref{fig:vp}. For this measurement we have assumed that the line gains $g_1$ and $g_3$ are not changed, while $g_4$ is obtained by repeating the procedure described before, i.e. by a  measurement of  $S{41} $, for each value of the pump power level.

The system noise temperature can be described by the following model

\begin{equation}
T_{\rm sys} \approx T_c +   \Lambda_1 T_{\rm TWPA}+  \frac{\Lambda_2 \Lambda_1}{G_{\rm TWPA}}T_{\rm HEMT}
\label{e:noise}
\end{equation}

where $T_c \simeq 100$ mK is the cavity temperature (the antenna is   critically coupled), $T_{\rm TWPA}$ and $G_{\rm TWPA}$ are the noise temperature and gain of the TWPA, $\Lambda_1$ are the insertion losses of the chain from point A1 to the TWPA input, $\Lambda_2$ are the insertion losses of the chain from the TWPA to the HEMT, and $T_{\rm HEMT}$ is the effective noise temperature of the HEMT. We have measured $T_{\rm HEMT} = 4.5 \pm 0.5$ K on a different set-up at the frequency of 10.5 GHz, while the two losses have been measured at room temperature, at least for the non superconducting cabling. An estimate for the losses at cryogenic temperature gives $\Lambda_1 = 0.3 $ dB and $\Lambda_2 = 0.7 $ dB. Such low losses allowed us to neglect noise contributions from the losses in Equation~(\ref{e:noise}). 
If we look at the plot of $T_{\rm sys}$ in Figure~\ref{fig:vp}, we see that there is no variation even in the presence of an increasing gain of the TWPA for pump levels above $- 24.8$ dBm. This indicates that the contribution of  $T_{\rm HEMT}$ is negligible.
If we consider the weighted average  of the system noise temperatures, 
excluding the one at the smallest pump value, we obtain $T_{\rm sys}^{\rm avg} = 3.3 \pm 0.1 $ K. We can then estimate $T_{\rm TWPA} = 3.0 \pm 0.1$ K at the operating frequency of 10.77 GHz. 

From Table \ref{t:results} follows that the output line has a total gain of $g_4$(dB)$\simeq 71.6$ dB. The nominal gains of the two HEMTs are 37 dB and 35 dB for the cryogenic and room temperature ones, respectively. We estimate a total lines loss of 20 dB, thus resulting in an estimated gain of the parametric amplifier $G_{\rm TWPA}\simeq 20$ dB. This value is compatible with the results showed in Figure \ref{fig:gain}, where the gain has been estimated only by switching the pump off. A precise measurement of $G_{\rm TWPA}$ is not in general necessary, since it is the total line gain which characterises the signal strength. A gain large enough to avoid contribution of the noise from the HEMT (see Equation~(\ref{e:noise})) is what is truly important.







\section{Conclusions}\
\label{Conclusioni}

We have operated and characterized an amplification chain to be used for the search of dark matter axions. It is based on a reversed Kerr travelling wave parametric amplifier and mounted to read the power delivered by a microwave resonant cavity.  The amplification chain will be the core of the experiment QUAX, designed to operate an axion haloscope in a wide bandwidth around 10 GHz \cite{2022arXiv220104223D}. A system noise temperature of $(3.3 \pm 0.1)$ K has been measured at a frequency of 10.77 GHz. This is a record low value for a wide band amplifier, and improves over commercially available HEMTs that have an internal noise contribution already above the value obtained here. A dedicated procedure has been devised to precisely measure the noise characteristic at the relevant measurement point, i.e. the cavity output.

\section{Acknowledgments}
We acknowledge Enrico Berto, Mario Tessaro and Fulvio Calaon for their work on the building and operation of the prototype,  and INFN - Laboratori Nazionali di Legnaro for hosting the experiment and providing the liquid Helium for cooling.

This work is supported by INFN (QUAX experiment) and by the European Union's Horizon
2020 research and innovation program under grant
agreement no. 899561. M.E. acknowledges the European
Union's Horizon 2020 research and innovation program
under the Marie Sklodowska Curie (grant agreement no.
MSCA-IF-835791). A.R. acknowledges the European
Union's Horizon 2020 research and innovation program
under the Marie Sklodowska Curie grant agreement No
754303 and the 'Investissements d'avenir' (ANR-15-
IDEX-02) programs of the French National Research
Agency.

The data that support the findings of this study are available from the corresponding author upon reasonable request.

\section*{References}

\bibliographystyle{unsrt}
\bibliography{TWC0p1}

\end{document}